\documentclass[twocolumn,preprintnumbers,amsmath,amssymb,prd]{revtex4}

\usepackage{epsfig}
\usepackage{graphicx}
\usepackage{dcolumn}
\usepackage{latexsym}
\usepackage{amssymb}

\newcommand{\be}{\begin{equation}}
\newcommand{\ee}{\end{equation}}
\newcommand{\bea}{\begin{eqnarray}}
\newcommand{\eea}{\end{eqnarray}}

\newcommand{\refc}[1]{(\ref{#1})}

\begin{document}

\title{Spectrum of the QCD flux tube in 3d SU(2) lattice gauge theory}
\author{Bastian B. Brandt}
\affiliation{Institut f\"ur Kernphysik, Johannes Gutenberg-Universit\"at Mainz, Johann Joachim Becher-Weg 45, D-55099 Mainz}
\email{brandt@kph.uni-mainz.de}
\author{Pushan Majumdar}
\affiliation{Dept. of Theoretical Physics, Indian Association for the Cultivation of Science, 
Jadavpur, Kolkata 700032.}
\email{tppm@iacs.res.in}

\preprint{MKPH-T-09-11}

\begin{abstract}
Evidence from the lattice suggests that formation of a flux tube between a $q\bar{q}$ pair in the
QCD vacuum leads to quark confinement. For large separations between the quarks, it is conjectured
that the flux tube has a behavior similar to an oscillating bosonic string, supported by lattice
data for the groundstate $q\bar{q}$ potential.
We measure the excited states of the flux tube in 3d SU(2) gauge theory with three different
couplings inside the scaling region. We compare our results to predictions of effective
string theories.
\end{abstract}

\maketitle

\section{Introduction}
Simulations over the last few years have accumulated strong evidence that gluonic dynamics 
 indeed leads to the formation of a flux tube between test 
quark and antiquark ($q\bar q$) in the vacuum of Yang-Mills theory \cite{bali,ymstring}.
This implies a linearly rising potential between quark and antiquark in the QCD vacuum and thus 
leads to quark confinement.
At large $q\bar q$ separations, this flux tube is expected to behave like a string. 
Open bosonic string descriptions of the dynamics of this flux tube have been 
attempted for a long time \cite{God}. 
Using the Nambu-Goto (NG) action, first \textit{Alvarez} \cite{Alvarez} (in the limit $d\to\infty$) and later
\textit{Arvis} \cite{Arvis} obtained the energy states of the flux tube as 
\begin{equation}
 \label{eqarvis}
E_{n}(R) = \sigma \: R \: \sqrt{ 1 + \frac{2\pi}{\sigma\:R^{2}} \: \left( n - \frac{1}
{24} \: ( d - 2 ) \right) }
\end{equation}
where $\sigma$ is the string tension, $R$ the quark-antiquark separation 
and $d$ the number of space-time dimensions.
A closed string description was proposed by Polchinski and Strominger (PS) \cite{PS}
 where they suggested how effective string theory with vanishing conformal 
anomaly could be formulated in arbitrary dimensions. 
The spectrum of the string using the PS prescription has been computed
in \cite{drum,pmd,unpub}. To order $R^{-3}$ it has been found that the spectrum 
is universal (depends only on the number of space-time dimensions and the string 
tension), and to this order it coincides with the NG spectrum.
For a calculation of the closed string spectrum on the lattice see \cite{teper}.
Using symmetry properties of the Polyakov loop correlation function, 
L\"uscher and Weisz (LW) showed that any effective string picture posseses 
an open-closed duality property \cite{lweisz}. They also showed that demanding 
this duality constrains the possible string spectra and in particular forbids 
$1/R^2$ terms in the effective string action. In the LW
formulation too, the spectrum is consistent with the NG spectrum. In 
3-dimensions it is exactly the same and in 4-dimensions one undetermined 
parameter remains which can of course assume the NG value. The NG partition 
function itself respects open-closed duality. Predictions from these theories 
can be tested by comparing them to results coming from simulations of pure 
 Yang-Mills theories.

In this Letter we will present a new scheme for measuring observables in simulations 
of pure Yang-Mills theories. Using this scheme we will accurately measure 
observables in 3-dimensional $SU(2)$ lattice gauge theory and try to extract the 
 excitation spectrum of the flux tube. For earlier studies of the 
QCD string spectra using different schemes in both 3- and 4-dimensions for 
several gauge theories, see \cite{Kuti} and the references therein.

\vspace*{-3mm}\section{Preliminaries}
To probe the properties of the flux tube formed between test quark and antiquark, 
the two observables in pure Yang-Mills theories 
are the Polyakov loop correlation function and spatio-temporal Wilson loops.
While Polyakov loop correlation functions project very strongly onto the ground state,
Wilson loops are much more sensitive to the excited states of the 
flux tube and in addition allows one to couple more strongly to a particular state 
while suppressing the others, by appropriate choice of the spatial parts of the 
loop. It is therefore the observable of choice if one wants to study the excitation 
spectrum of the flux tube. 

\vspace*{-3mm}\subsection{Extraction of the excited states}
\begin{figure}[htb]
\centering
\includegraphics[width=.45\textwidth]{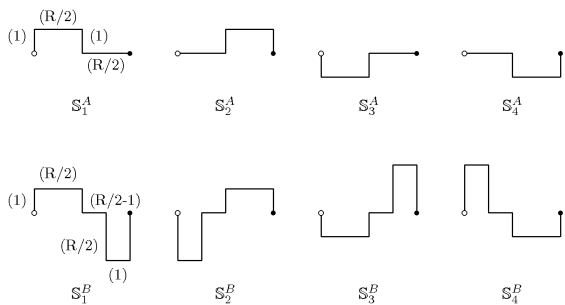}
\caption{\textbf{Top:} Basic set of operators used to construct $(C,P)$ channels.
\textbf{Bottom:} Set of operators with stronger coupling to the second excited state.}
\label{fig1}
\end{figure}

In 3-dimensions, the states of the 
oscillating string can be classified by charge-conjugation and parity properties $(C,P)$ 
\cite{Pushan2,dipl}. Choosing the spatial parts of the Wilson loops  
as shown in figure \ref{fig1}, the $(C,P)$ projectors are given by the 
superpositions 
\begin{equation}
 \label{eq2.8b}
\begin{array}{rcl}
 \mathbb{S}^{++} & = & \mathbb{S}_{1} + \mathbb{S}_{2} + \mathbb{S}_{3} + \mathbb{S}_{4} \vspace*{2mm} \\
 \mathbb{S}^{+-} & = & \mathbb{S}_{1} + \mathbb{S}_{2} - \mathbb{S}_{3} - \mathbb{S}_{4} \vspace*{2mm} \\
 \mathbb{S}^{--} & = & \mathbb{S}_{1} - \mathbb{S}_{2} - \mathbb{S}_{3} + \mathbb{S}_{4} \vspace*{2mm} \\
 \mathbb{S}^{-+} & = & - \mathbb{S}_{1} + \mathbb{S}_{2} - \mathbb{S}_{3} + \mathbb{S}_{4} 
\end{array}
\end{equation}
where $+/-$ stands for even or odd states under $C$ and $P$.
Although in each of the $(C,P)$ channels, there are an infinite number of states, we are 
going to look only at the ground states and will label these channels as $\{0,1,2,3\}$ 
respectively. The Wilson loop projecting onto a channel $n$ has the spectral 
representation 
\begin{equation}
 \label{eq2.9}
W_n(R,T) = \sum_{i=0}^{\infty} \beta_i(R)\: e^{ - E_n^i(R) \: T }.
\end{equation} 
where $E^i$ are the energies of the states in a given channel.
Using Wilson loops with different temporal extents, one obtains 
 the energies and energy differences at leading order as
\begin{widetext}
\begin{eqnarray}
 \label{eq2.12}
- \frac{1}{T_{b}-T_{a}} \: \ln \left[ \frac { W_{n} ( R , T_{b} ) } { W_{n} ( R , T_{a} ) } \right] & = & E_{n} ( R ) + \frac{1}{T_{b}-T_{a}} \: \alpha_{n}(R) \: e^{ - \delta_{n}(R) \: T_{a} } \: \left( 1 - e^{ - \delta_{n}(R) \: ( T_{b} - T_{a} ) } \right) \\
 \label{eq2.13}
- \frac{1}{T_{b}-T_{a}} \: \ln \left[ \frac { W_{n} ( R , T_{b} ) \: W_{m} ( R , T_{a} ) } { W_{n} ( R , T_{a} ) \: W_{m} ( R , T_{b} ) } \right] & = & \Delta E_{nm}(R) + \frac{1}{T_{b}-T_{a}} \: \alpha_{n}(R) \: e^{ - \delta_{n}(R) \: T_{a} } \: \left( 1 - e^{ - \delta_{n}(R) \: ( T_{b} - T_{a} ) } \right)
\end{eqnarray}
\end{widetext}
where $T_{a}<T_{b}$. 
In the following, the values obtained from the LHS of eqns.\refc{eq2.12} and \refc{eq2.13}
 will be 
called na\"ive values and be denoted by $\bar{E}_{n}(R)$ and $\Delta \bar{E}_{nm}(R)$.
The quantities $E_{n}(R)$ and $\Delta E_{nm}(R)$ will be called corrected values. 
They, along with $\alpha_{n}$ and $\delta_{n}$, are obtained as fit parameters. The fits 
 are done by taking all possible combinations of $T_a$ and $T_b$ and are discussed 
in more detail in Section D.
 
\vspace*{-2mm}\subsection{Noise reduction techniques}
String like behavior of the flux tube is expected to 
occur at large $q\bar q$ separations. The Wilson loops we measure must therefore extend to large 
enough $R$ values. Also we see from equations \refc{eq2.12} and \refc{eq2.13} that to 
reduce the contaminations due to other states we must either go to large values of $T$ 
or tune $\alpha$ to small values by appropriate choice of the basis states. The latter method 
has been followed in \cite{Juge} where asymmetric lattices with small temporal extents 
were used. The former method requires accurate measurements of expectation values of large Wilson loops,
made possible recently by the multilevel algorithm, proposed in 
\cite{Weisz1}. This method has been 
used in \cite{Weisz2,PundD} to accurately measure the ground state properties of the 
flux tube. Here we will try to combine both methods by using  a slight variant of the 
multilevel algorithm.

\begin{figure}
\centering
\includegraphics[width=.2\textwidth]{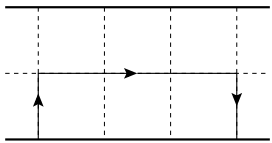}
\caption{The source of the Wilson loop in the modified algorithm. The product of the directed lines 
defines the source. The thick lines are held fixed during sub-lattice updates.}
\label{staplefig}
\end{figure}

In the multilevel algorithm, intermediate averages are computed for the temporal links by 
updating certain sub-lattices with the sources on the spatial links of the boundaries 
of these sub-lattices. The boundaries are held fixed during the sub-lattice updates.
We now put the sources on space-like surfaces in the middle of a sub-lattice of 
thickness $2a$ where $a$ is the lattice spacing. Two temporal links are attached to the two 
ends of the source which 
terminates on the space-like surfaces that are fixed during the sub-lattice updates 
(see fig.\ref{staplefig}). The advantage is that the sub-lattice updates reduces 
fluctuations of the sources as well as fluctuations of the temporal 
links. Moreover, in addition to using multihit on the temporal links, we can now use 
multihit on the spatial links as well. For further details see \cite{ownPos,dipl} 
and references therein. 
Adopting the notation in \cite{Weisz1}, we now define the expectation value of the 
Wilson loop by 
\begin{equation}
\langle W(T) \rangle = \langle \{\mathbb{L}(0)\}_{\alpha\gamma}\{\mathbb{T}(a)\cdots \mathbb{T}((T-1)a)\}_{\alpha\beta\gamma\delta}
\{\mathbb{L}(T)^{\ast}\}_{\beta\delta} \rangle ,
\end{equation}
where $\mathbb{L}(t)$ is the operator of fig.\ref{staplefig} and $\{~\}$ indicates sub-lattice 
averaged quantities.
The only difference from \cite{Weisz1} is that now we have $\{\mathbb{L}\}$ instead of $\mathbb{L}$.
A similar method was used in \cite{forcrand} to measure the static potential and observe string breaking in the adjoint representation.

\begin{figure}
\centering
\begin{minipage}[c]{0.45\textwidth}
\includegraphics[angle=-90, width=\textwidth]{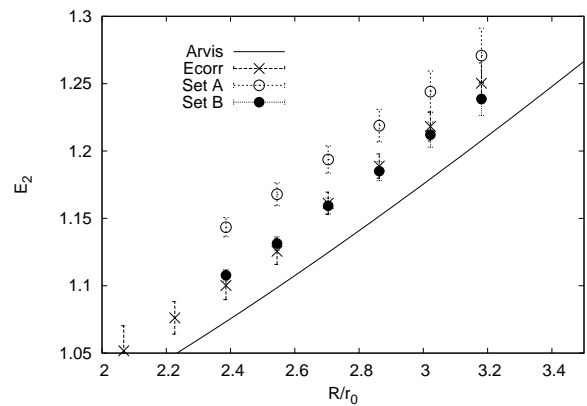}
\caption{Comparison of the na\"ive energy $\bar{E}_{2}$ from set $A$ and $B$.}
\label{imprsorfig}
\end{minipage}
\end{figure}

In most parts of the calculations, the operators
$\left\{\mathbb{S}^{A}_{i}\right\}$ (figure \ref{fig1} top; see also \cite{Pushan2,ownPos}) are good enough for a reliable signal.
However, for the second excited state beyond $\beta=7.5$ the error reduction was not
sufficient and we had to use 
another set of operators $\left\{\mathbb{S}^{B}_{i}\right\}$,
(figure \ref{fig1} bottom) with a stronger coupling to that particular excited state.
We will henceforth refer to the two sets as set $A$ and set $B$, respectively.
To check the error reduction with the new set of operators we performed 460 measurements at $\beta=7.5$
with spatial extents $R=15-20$ using the simulation parameters shown in table \ref{tab2}.
The resulting relative errors of $\bar{E}_{2}$ are also tabulated there. The values for the
na\"ive energies using both sets are plotted in figure \ref{imprsorfig}. For comparison we also 
plot $E_2$, which was calculated from $\bar E_2^A$, in the same region.
It is clearly seen that not only do the na\"ive values of set $B$ already coincide 
with the corrected values of set $A$, they also have much smaller error bars.
Set $B$ therefore significantly improves the signal for the state $\mathbb{S}^{--}$.

There are several parameters for the algorithm, that have to be tuned properly in order to achieve 
the maximal error reduction.
For the temporal links we follow the same procedure as outlined in \cite{Pushan2}. In addition we 
now have another parameter which is the number of updates ($N_{s}$) for the sub-lattices 
containing the spatial operators. We found it beneficial for the excited states 
to have quite a large number of such sub-lattice updates. Since there are only two such 
sub-lattices
 for every loop this does not increase the cost of the simulation very much. Table \ref{tab1} 
contains the resulting optimized run parameters and we refer to \cite{dipl}
for details of the optimization. 

\begin{table}[b]
\centering
\begin{tabular}{c|cccccc|cccccc}
$T$ & Lat & $t_{s}$ & $N_{s}$ & $N_{t}$ & \hspace*{3mm} & $R$ & 15 & 16 & 17 & 18 & 19 & 20 \\
\hline \hline
6 & 36$^{3}$ & 4 & 18000 & 1500 & & $A$ & 0.62 & 0.73 & 0.85 & 0.99 & 1.23 & 1.60 \\
10 & 40$^{3}$ & 4 & 18000 & 2500 & & $B$ & 0.39 & 0.45 & 0.50 & 0.60 & 0.77 & 0.99
\end{tabular}
\caption{\textbf{Left:} Parameters of the testruns to compare the operators. \textbf{Right:} Relative error of $\bar{E}_{2}$ in \% for operator sets $A$ and $B$.}
\label{tab2}
\end{table}

\vspace*{-2mm}\subsection{Simulation parameters and lattice scales}
\begin{table*}
\centering
\begin{tabular}{l|ccc|cc|ccccccl}
Lat & $\beta$ & $r_{0}$ & $ a\left[fm\right]$ & $R$ & $\left\{S_{i}\right\}$ &
$T$ & $T\left[fm\right]$ & $t_{s}$ & size & $N_{s}$ & $N_{t}$ & \#meas \\
\hline
\hline
 & & & & & & & & & & & & \vspace*{-2mm} \\
$L^{A}$ & 7.5 & 6.2875(10) & 0.07952(1) & $7-20$ & A & 6 & 0.477 & 4 & $38^{3}$ & 36000 & 1500 & 4400 \\
 & & & & & & 10 & 0.795 & & $40^{3}$ & & 3000 & 6468 \\
 & & & & & & 14 & 1.113 & & $42^{3}$ & & 9000 & 11176 \\
 & & & & & & 18 & 1.431 & & $54^{3}$ & & 18000 & 6512 \\
\hline
 & & & & & & & & & & & & \vspace*{-2mm} \\
$L_{A}^{B}$ & 10.0 & 8.6602(8) & 0.05812(1) & $9-27$ & A & 8 & 0.465 & 6 & $40^{3}$ & 48000 & 3000 & 1272, 1272, 1296$^{\ast}$ \\
 & & & & & & 10 & 0.581 & 4 & $50^{3}$ & & 3000 & 2352, 2544, 2568 \\
 & & & & & & 14 & 0.814 & 6 & $56^{3}$ & & 6000 & 6384, 6216, 6480 \\
 & & & & & & 18 & 1.046 & 4 & $54^{3}$ & & 12000 & 8664, 8304, 8472 \\
\hline
 & & & & & & & & & & & & \vspace*{-2mm} \\
$L_{B}^{B}$ & 10.0 & 8.6602(8) & 0.05812(1) & $9-27$ & B & 6 & 0.349 & 4 & $48^{3}$ & 48000 & 1500 & 2000, 2000$^{\ast\ast}$ \\
 & & & & & & 8 & 0.465 & 6 & $48^{3}$ & & 3000 & 2000, 6000 \\
 & & & & & & 10 & 0.581 & 4 & $50^{3}$ & & 6000 & 2000, 7960 \\
 & & & & & & 14 & 0.814 & 6 & $56^{3}$ & & 12000 & 2000, 2000 \\
\hline
 & & & & & & & & & & & & \vspace*{-2mm} \\
$L^{C}$ & 12.5 & 10.916(3) & 0.04580(1) & $11-29$ & A\&B & 8 & 0.366 & 6 & $48^{3}$ & 36000 & 2000 & 1000 \\
 & & & & & & 10 & 0.458 & 4 & $50^{3}$ & & 3000 & 4000 \\
 & & & & & & 14 & 0.641 & 6 & $56^{3}$ & & 6000 & 7080 \\
 & & & & & & 18 & 0.824 & 4 & $72^{3}$ & & 12000 & 2080 \\
\hline
\hline
\end{tabular}
\caption{Run parameters of the simulations. The number of measurements marked with $^{\ast}$ corresponds to the $R$ values 
$9-15,17-21,23-27$ and the ones marked with $^{\ast\ast}$ to $9-15,17-27$ respectively.}
\label{tab1}
\end{table*}
Our simulations were done with three different couplings, $\beta=7.5,10.0$ and $12.5$ 
(chosen so as to lie in the scaling region) in
3-dimensional $SU(2)$ lattice gauge theory, using usual heatbath sweeps \cite{heatb},
combined with three overrelaxation sweeps.
The scale was set by the Sommer parameter $r_{0}$ \cite{sommer}, which has been computed for these
$\beta$ values with high accuracy in \cite{PundD}.
For each beta value we calculated Wilson loops with four different temporal extents and used cubic
lattices. 
 
Our simulation parameters, lattice scales, along with the parameters for the multilevel algorithm are
also tabulated in table \ref{tab1}.

\vspace*{-2mm}\subsection{Error analysis and control of the fits}

For estimating the error of all the energy values and differences, we used the usual binned 
jackknife method, with 44, 24, 40 and 40 bins for the lattices $L^A$, $L^B_A$, $L_B^B$ 
and $L^C$, respectively. We also checked that the errors did not vary by more than 
a few percent with bin size. 

For the energy difference $\Delta E_{20}$ at $\beta=10.0$ and $12.5$ we used the 
results obtained from set $B$ for $E_{2}$ while the
values for $E_{0}$ were obtained from set $A$. Since these two values come from 
independent simulations, we
added the individual errors in quadrature to obtain error estimates for $\Delta E_{20}$.

The remaining issue is the control of the fits \refc{eq2.12} and \refc{eq2.13},
and this was done in two ways.
For the energies $E_{n}$, we expect $\alpha$
to be smaller then the ratio of the degeneracies \footnote{At the order where we are 
comparing the data with the theory, the degeneracies are the same as in the free theory 
\cite{Weisz2}} of the energy states considered and $\delta$ should be of the order
of the energy gap to the next level in the channel.
Similar conclusions hold for the parameters of the energy differences \cite{Pushan2,dipl}
 as well.
If the resulting fit parameters were far of from these expectations we did not trust 
the fits.
As a second check we plot the expected corrections $$\Delta= 
\frac{1}{T_{b}-T_{a}} \: \alpha_{n}(R) \: e^{ - \delta_{n}(R) \: T_{a} } \: 
\left( 1 - e^{ - \delta_{n}(R) \: ( T_{b} - T_{a} ) } \right),$$ obtained with 
averaged parameters $\alpha_{n}$ and $\delta_{n}$
against the differences $\bar{\Delta}=E_{n}-\bar{E}_{n}$, for all possible combinations
of $T_{a}$ and $T_{b}$. We show an example for these fits ($\Delta E_{10}$, $\beta=7.5$, $R=11$
and $15$) in figure \ref{check1fig}.
If for some of the values there was a big discrepancy to the expectation 
$\Delta=\bar{\Delta}$ the fit was also regarded
as unreliable (for more details and systematics of the fits see \cite{dipl} and \cite{Pushan2}).

\vspace*{-2mm}\subsection{Finite volume effects}

Finite volume effects are possible due to finite spatio-temporal extents of the 
lattices. To control the effects due to finite temporal extents, we use four different 
extents to make sure that any systematic error in our energy values is much lower than 
our statistical errors.

The dominant corrections due to finite spatial extents are around-the-world glueball exchanges
of the form $a(R)\:\exp (-m_G(L-R))$ , where 
$m_G$ is the glueball mass and $L$ is the spatial extent of the lattice. These are 
relevant for large values of $R$. While glueball masses for 
3-dimensional SU(2) lattice gauge theory has been studied in \cite{cmichael,miketeper}, 
the mixing coefficient $a(R)$ is not known. To get an estimate for this coefficient, 
we did runs at different lattice volumes at $\beta=5$ (for which $m_G$ has been measured in 
\cite{miketeper}), varying $L$ by a factor two. Since we did not see any finite size effects in the energies, we concluded, that in the range
considered, $a(R)$ is a number of order one.

Using interpolation, we obtain from the data in \cite{miketeper}, the glueball masses at $\beta$ = 
7.5, 10.0 and 12.5 to be 0.856, 0.671 and 0.536 respectively in lattice units. Since our data at 
large $R$, with systematic errors under control is only at 
$\beta = 7.5$, we concentrate on that $\beta$ value. The only lattice where 
this correction matters is the $L=42$ lattice on which we measure Wilson loops with
$T=14$ and $R$ between $7-20$.
On this lattice, the correction ranges between $10^{-9}$ and $6.7\times 10^{-9}$ for 
$R$ between $18-20$. The error bars for the wilson loop corresponding to the second excited state are of 
similar magnitude for $R=19$ and 20 while for the first excited state, $R=20$ has a similar error bar.
We therefore expect that there can be some correction due to the glueball for these energy values while 
for all other values, the corrections are well below our statistical errors.
 
\vspace*{-2mm}\section{Simulation Results}
\label{results}

In this section we discuss the results of our simulations, that are tabulated in table \ref{tab3}.
We compare to the full NG spectrum, eq.\refc{eqarvis} and to leading order (LO) and 
next to leading order (NLO) models which are obtained by truncating the expansion of the 
square root of \refc{eqarvis} in $1/R^{2}$ at leading order and next to leading order 
respectively. The LW and PS type string theories give identical results to NLO order.
The curves in the plots were drawn using the string tension at $\beta=12.5$.  All the 
results have been rescaled so that they are visible on a single plot.

\vspace*{-2mm}\subsection{Energy states}

\begin{figure}
\centering
\begin{minipage}[c]{0.45\textwidth}
\includegraphics[width=\textwidth]{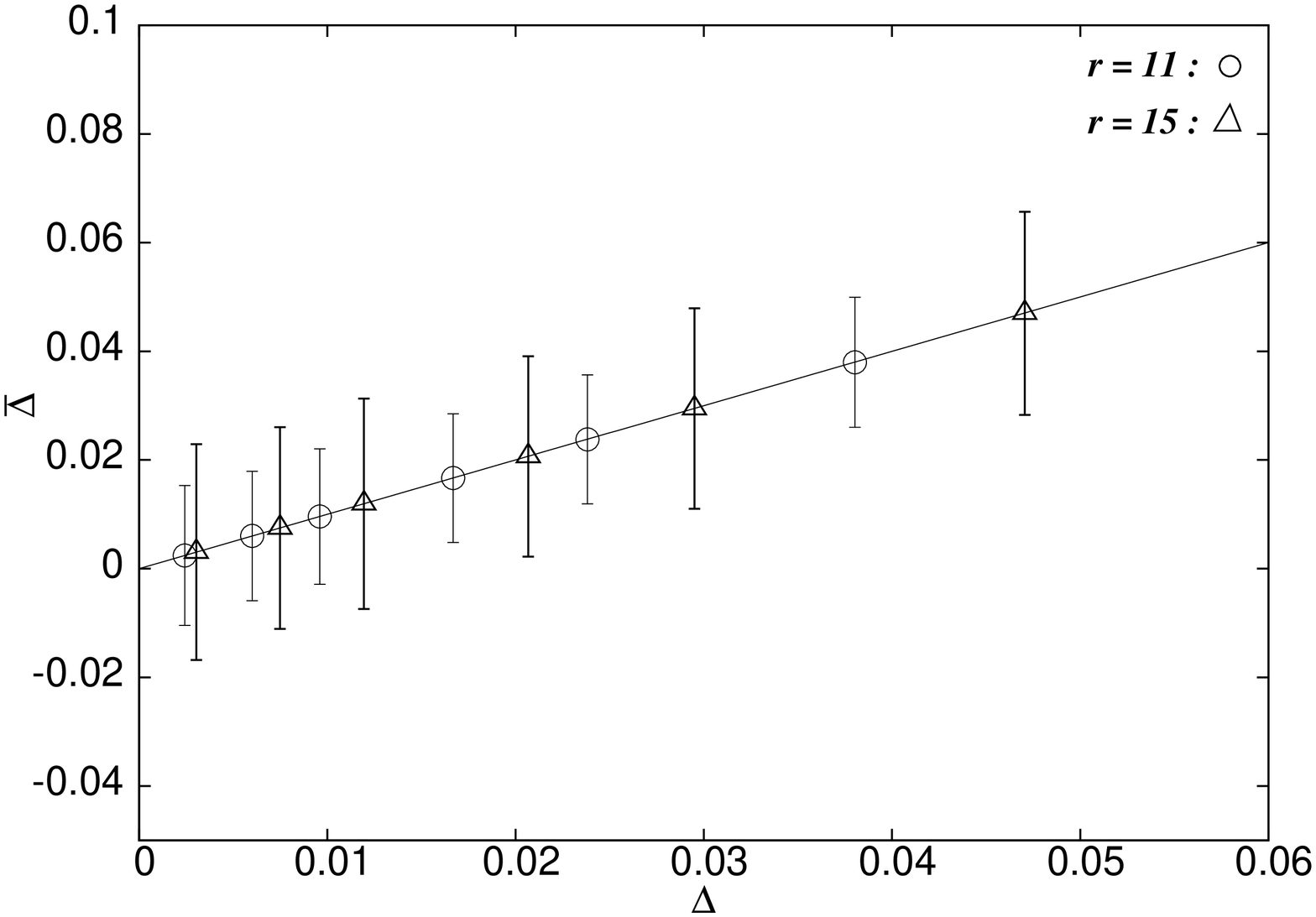}
\end{minipage}
\caption{Cross check of the fits; $\Delta E_{10},\, \beta =7.5$.}
\label{check1fig}
\begin{minipage}[c]{0.5\textwidth}
\includegraphics[angle=-90, width=\textwidth]{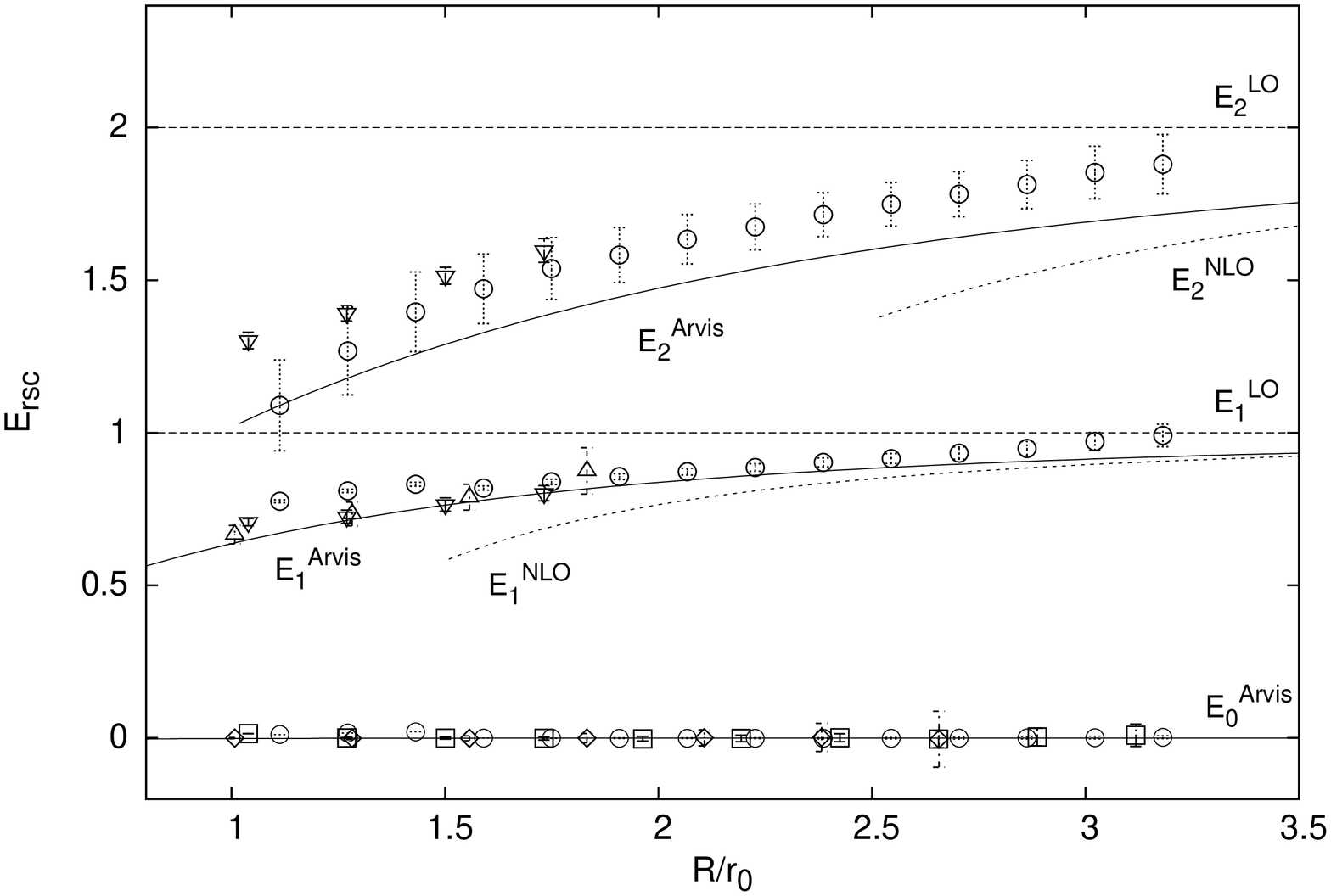}
\end{minipage}
\caption{Results for the total energies $E_{n}$. The $\bigcirc$'s are the values for 
$\beta=7.5$, $\Box$'s for $\beta=10.0$ and $\Diamond$'s for $\beta=12.5$. 
 $\bigtriangledown$ are corrected values for $\beta=10.0$ and 
 $\bigtriangleup$ for $\beta=12.5$. 
The results have been rescaled, such that $E_{n}^{\textnormal{LO}}=n$. 
The labeling of the lines are as defined in the text.
}
\label{fig2}
\end{figure}

\begin{table*}[t]
\centering
\small
\begin{tabular}{c|ll|ll|ll|lc|c}
\hline
\multicolumn{1}{c|}{$R$} & \multicolumn{2}{c|}{$E_{0}$} & \multicolumn{2}{c|}{$E_{1}$} & \multicolumn{2}{c|}{$E_{2}$} & 
\multicolumn{2}{c|}{$\Delta E_{10}$} & $\Delta E_{20}$ \\
\hline
\hline
 & & & & & & & & & \vspace*{-2mm} \\
$\beta=7.5$ & \multicolumn{1}{c}{$T(18,14)$} & \multicolumn{1}{c|}{$4T\,{\mathbf \ast}$} & \multicolumn{1}{c}{$T(18,14)$} & 
\multicolumn{1}{c|}{$4T\,{\mathbf \ast}$} & \multicolumn{1}{c}{$T(18,14)\,{\mathbf \ast}$} & \multicolumn{1}{c|}{} & 
\multicolumn{1}{c}{$T(18,14)$} & \multicolumn{1}{c|}{$4T\,{\mathbf \ast}$} & \multicolumn{1}{c}{$T(18,14)\,{\mathbf \ast}$} \\
\hline
 & & & & & & & & & \vspace*{-2mm} \\
7 & 0.4251(1) & 0.42992(8) & 0.753(1) & 0.773(2) & 0.91(7) & & 0.328(1) & 0.302(3) & 0.49(7) \\
8 & 0.4660(1) & 0.47226(9) & 0.768(1) & 0.784(2) & 0.96(6) & & 0.302(1) & 0.284(2) & 0.50(6) \\
9  & 0.5064(2) & 0.5135(1) & 0.785(1) & 0.797(3) & 0.99(5) & & 0.279(1) & 0.267(3) & 0.49(5) \\
10 & 0.5464(2) & 0.5464(2) & 0.806(1) & 0.804(2) & 1.01(4) & & 0.259(1) & 0.257(3) & 0.46(4) \\
11 & 0.5863(2) & 0.5862(3) & 0.828(1) & 0.826(2) & 1.03(3) & & 0.242(1) & 0.240(3) & 0.43(7) \\
12 & 0.6259(3) & 0.6259(3) & 0.853(1) & 0.850(2) & 1.04(2) & & 0.227(1) & 0.224(3) & 0.41(2) \\
13 & 0.6655(3) & 0.6654(4) & 0.879(1) & 0.876(3) & 1.06(2) & & 0.214(1) & 0.211(3) & 0.40(2) \\
14 & 0.7049(3) & 0.7047(5) & 0.907(1) & 0.904(3) & 1.08(2) & & 0.202(1) & 0.199(3) & 0.38(2) \\
15 & 0.7442(4) & 0.7441(5) & 0.936(1) & 0.933(3) & 1.10(2) & & 0.192(1) & 0.189(3) & 0.36(2) \\
16 & 0.7835(4) & 0.7833(6) & 0.966(2) & 0.963(3) & 1.13(1) & & 0.183(2) & 0.180(4) & 0.34(1) \\
17 & 0.8228(5) & 0.8225(7) & 0.998(2) & 0.995(4) & 1.15(1) & & 0.175(2) & 0.172(4) & 0.33(1) \\
18 & 0.8620(5) & 0.8617(7) & 1.030(2) & 1.027(4) & 1.18(1) & & 0.168(2) & 0.165(4) & 0.32(1) \\
19 & 0.9012(5) & 0.9009(8) & 1.064(3) & 1.061(5) & 1.21(1) & & 0.162(3) & 0.160(5) & 0.31(1) \\
20 & 0.9404(6) & 0.9401(8) & 1.097(4) & 1.095(6) & 1.24(2) & & 0.157(4) & 0.155(6) & 0.29(2) \\
 & & & & & & & & & \vspace*{-2mm} \\
\hline
 & & & & & & & & & \vspace*{-2mm} \\
$\beta=10.0$ & \multicolumn{1}{c}{$T(18,14)$} & \multicolumn{1}{c|}{$4T\,{\mathbf \ast}$} & \multicolumn{1}{c}{$T(18,14)$} & 
\multicolumn{1}{c|}{$3T\,{\mathbf \ast}$} & \multicolumn{1}{c}{$T(10,8)$} & \multicolumn{1}{c|}{$3T\,{\mathbf \ast}$} & 
\multicolumn{1}{c}{$T(18,14)$} & \multicolumn{1}{c|}{} & \multicolumn{1}{c}{} \\
\hline
 & & & & & & & & & \vspace*{-2mm} \\
9  & 0.3160(1) & 0.3207(1) & 0.571(1) & 0.563(4) & 0.772(4) & 0.770(9) & 0.255(1) & Best &  Best\\
11 & 0.3599(1) & 0.3596(5) & 0.587(1) & 0.567(6) & 0.770(3) & 0.757(7) & 0.227(2) & estimate & estimate \\
13 & 0.4032(2) & 0.4026(8) & 0.609(2) & 0.588(5) & 0.780(3) & 0.769(7) & 0.206(2) & is & is \\
15 & 0.4460(2) & 0.445(1)  & 0.634(2) & 0.613(5) & 0.794(4) & 0.780(8) & 0.188(2) & $E_{1}-E_{0}$ & $E_{2}-E_{0}$  \\
17 & 0.4884(3) & 0.487(1)  & 0.663(3) & & 0.809(2) & & 0.175(3) & &  \\
19 & 0.5309(3) & 0.529(2)  & 0.694(3) & & 0.834(3) & & 0.163(3) & &  \\
21 & 0.5733(4) & 0.571(2)  & 0.726(4) & & 0.860(3) & & 0.152(4) & &  \\
23 & 0.6153(5) & 0.613(4)  & 0.757(7) & & 0.889(4) & & 0.142(7) & &  \\
25 & 0.6575(8) & 0.655(4)  & 0.79(1)  & & 0.919(6) & & 0.14(1)  & &  \\
27 & 0.699(2)  & 0.697(4)  & 0.84(3)  & & 0.954(10)& & 0.14(3)  & &  \\
 & & & & & & & & & \vspace*{-2mm} \\
\hline
 & & & & & & & & & \vspace*{-2mm} \\
$\beta=12.5$ & \multicolumn{1}{c}{$T(18,14)$} & \multicolumn{1}{c|}{$4T\,{\mathbf \ast}$} & \multicolumn{1}{c}{$T(18,10)$} & 
\multicolumn{1}{c|}{$3T\,{\mathbf \ast}$} & \multicolumn{1}{c}{$T(14,10)$} & \multicolumn{1}{c|}{} & \multicolumn{1}{c}{$T(18,10)$} & 
\multicolumn{1}{c|}{} & \multicolumn{1}{c}{} \\
\hline
 & & & & & & & & & \vspace*{-2mm} \\
11 & 0.2540(2) & 0.2535(5) & 0.472(1) & 0.444(9)   & 0.612(3) & & 0.218(1) & Best &  Best\\				      
14 & 0.2953(3) & 0.2945(9) & 0.488(2) & 0.459(9)   & 0.613(3) & & 0.192(2) & estimate & estimate \\			      
17 & 0.3359(4) & 0.335(1)  & 0.510(2) & 0.481(8)   & 0.625(3) & & 0.173(2) & is & is \\ 				      
20 & 0.3761(6) & 0.374(2)  & 0.536(3) & 0.512(12)  & 0.644(4) & & 0.158(3) & $E_{1}-E_{0}$ & $E_{2}-E_{0}$  \\
23 & 0.4162(9) & 0.414(4)  & 0.566(5) & & 0.663(5) & & 0.147(5) & &  \\
26 & 0.456(1)  & 0.453(6)  & 0.593(9) & & 0.688(7) & & 0.134(9) & &  \\
29 & 0.496(2)  & 0.492(10) & 0.622(16)& & 0.707(11)& & 0.122(16)& &  \\
\hline
\end{tabular}
\normalsize
\caption{Results of the simulations. The data columns used in the plots have been denoted by a $\ast$ on their 
headers. The left column corresponds to the na\"ive value of loops, obtained from 
Wilson loops with the temporal extents given in brackets in the header. 
The right column corresponds to the 
 value from a fit to the form \refc{eq2.12},\refc{eq2.13}, wherever this was possible and the resulting
values are expected to have negligible systematic errors. The header $4T$ 
means, that the values were obtained using fits to all possible combinations of the four different 
temporal extents, and the header $3T$ means that combinations from the three lowest temporal extents were used.
Where these fits did not work and the data may contain significant systematic errors, we have only indicated what our 
best estimate is. The na\"ive data serves both for future referencing as well as estimating the corrections due to finite 
temporal extents. They give useful upper bounds.}
\label{tab3}
\end{table*}

\begin{table*}[b]
\centering
\small
\begin{tabular}{c|cc|ccc|ccc}
\hline
 & & & & & & & & \vspace*{-2mm} \\
 $\beta$ & 7.5 set $A$ & \cite{PundD} & 10.0 set $A$ & 10.0 set $B$ & \cite{PundD} & 12.5 set $A$ & 12.5 set $B$ & \cite{PundD} \\
\hline
\hline
 & & & & & & & & \vspace*{-2mm} \\
$ \sigma $ & 0.03867(7) & 0.038566(6) & 0.0206(2) & 0.0210(4) & 0.020606(4) & 0.0128(4) & 0.0129(4) & 0.012742(17) \\
$ V_{0} $ & 0.1730(5) & & 0.145(3) & 0.139(6) & & 0.124(4) & 0.124(4) & \\
\hline
\end{tabular}
\caption{Results for the string tension $\sigma$ and the regularization constant $V_{0}$. The values \cite{PundD} are 
the reference results for the string tension.}
\label{tab4}
\end{table*}

We use the groundstate to determine the string tension and fix an additive
constant $V_{0}$, appearing in the potential, by fitting to the form \cite{Arvis}:
\begin{equation}
\label{res-1}
V(R) = \sigma \: R \: \sqrt{ 1 - \frac{\pi}{12\:\sigma\:R^{2}} } + V_{0}
\end{equation}
The results of the fits are shown in table \ref{tab4}, where also the values from 
\cite{PundD} are shown
for comparison. We see that all results agree well within error bars. 
Also $\sigma$ and $V_{0}$ from both operator sets agree with each other.

In figure \ref{fig2} the results for the total energies are shown against $R/r_{0}$,
together with the predictions of the NG spectrum. 
The energies are rescaled such that $E_{n}^{LO}=n$.
The results for the groundstate are completely in agreement with
the NG predictions. In this case only the curve of the full spectrum is shown since deviations
at different orders of the expansion are not visible on this scale.

For the excited states we were only able to obtain corrected results over a limited range 
of $R$, except for $E_{1}$ at $\beta=7.5$ where we have results over the full range.

We see that the corrected values for $\beta=7.5$ follow the Nambu curves quite well. 
In the region where we are able to obtain corrected energies for all values of $\beta$,
we see good agreement between the different $\beta$ values. 
At smaller values of $R$ where the difference between the NG, LO and NLO curves are 
visible, the data seems to favor the NG curve.

\vspace*{-2mm}\subsection{Energy differences}

\begin{figure}[t]
\centering
\includegraphics[angle=0, width=.5\textwidth]{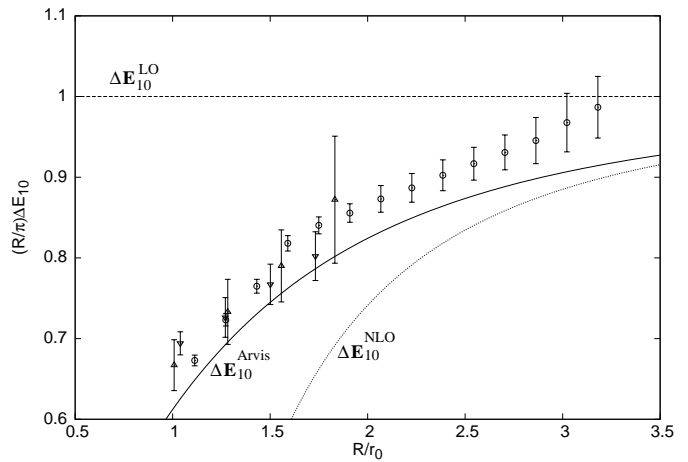}
\includegraphics[angle=0, width=.5\textwidth]{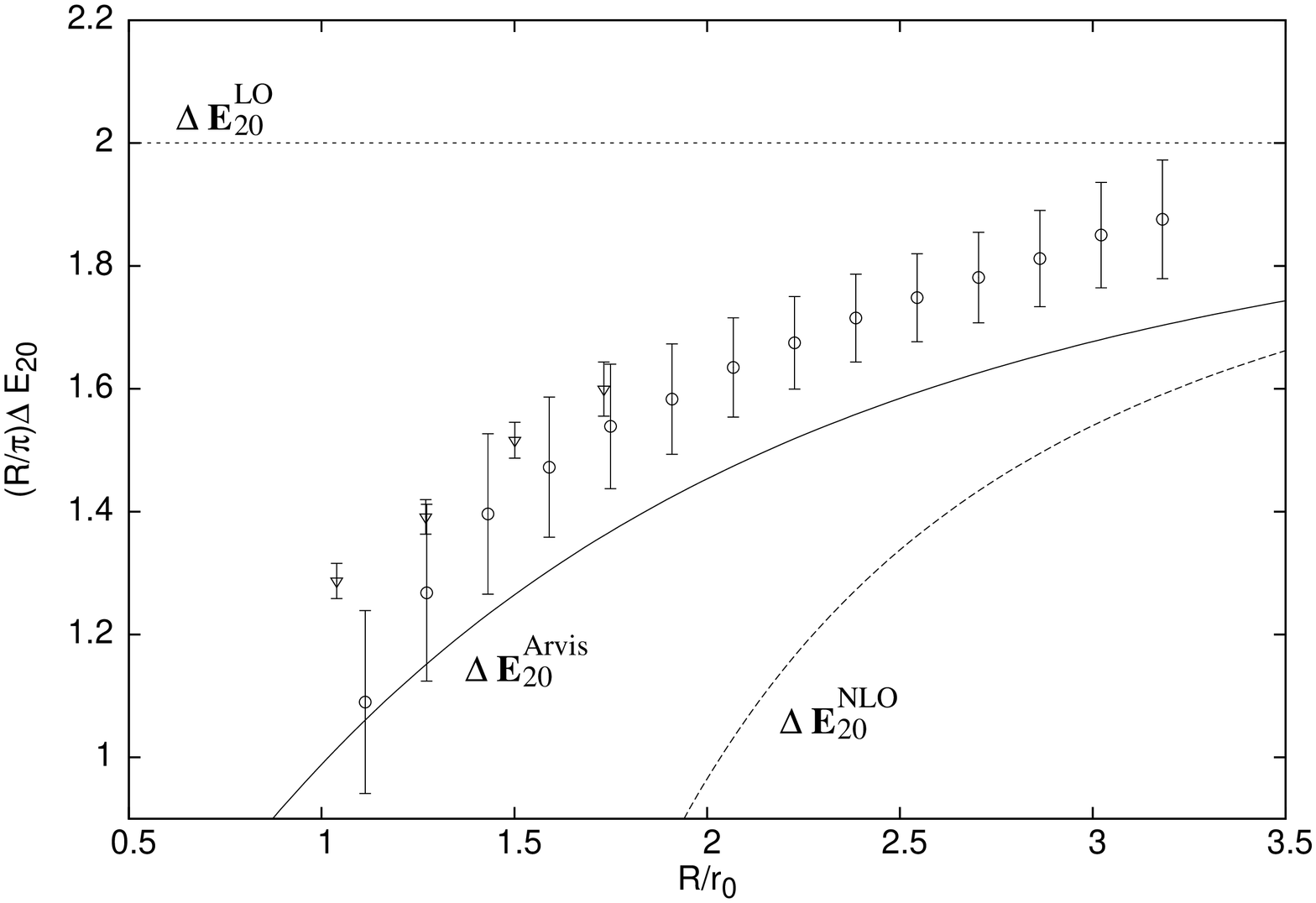}
\caption{\textbf{Top:} Results for the energy difference $\Delta E_{10}$. \textbf{Bottom:} Results for the 
difference $\Delta E_{20}$. The mapping of the curves and the points is the same as in the plot for the total energies.}
\label{fig3}
\end{figure}

We now turn to the energy differences which are more sensitive to subleading properties of the
flux tube. In addition they have the advantage that the constant $V_{0}$ does not contribute.
The results for the energy differences $\Delta E_{10}$ and $\Delta E_{20}$ are shown in figure \ref{fig3}.

We were able to obtain corrected values using eq.\refc{eq2.13} only for $\Delta E_{10}$ 
at $\beta=7.5$. That data set seems to follow the NG curves quite nicely. For $\Delta E_{20}$, our 
best estimates are the na\"ive values from eq.\refc{eq2.13} with $T_a=14$ and $T_b=18$. 
Nevertheless we expect very little systematic effects in these results
as the physical temporal extents for these loops are $>$ 1fm.
Again we see that the NG curve is favored by the data.

For $\Delta E_{10}$ at $\beta=10.0$ and 12.5, our best estimates come from the difference $E_1-E_0$ 
with the errors being calculated in quadrature. Even though this gives larger error bars, especially 
at $\beta=12.5$, we see that both data sets are consistent with $\beta=7.5$ values.

For $\Delta E_{20}$ at $\beta=10.0$ and 12.5, the physical temporal extents of the loops are not 
large enough that the higher order corrections, unaccounted for in equations \refc{eq2.12} and 
\refc{eq2.13}, are negligible. Nevertheless, at $\beta=10.0$, where it was possible at least 
partly to take into account the corrections, we see the trend of the data is to follow the NG curve.
For $\beta=12.5$, where even that was not possible, the data contains systematic errors and we 
have not plotted it. 

\vspace*{-2mm}\section{Conclusions}

In this Letter we have looked at a variant of the multi-level error reduction scheme suitable for studying 
the excited states of the QCD flux tube. Using this scheme 
we have looked at the excited states of the flux tube at three different couplings for 
$q\bar{q}$ separations between 0.5 and 1.7 fm. 

Compared to measurements with the older method \cite{Pushan2}, we have seen very significant improvements due to the 
new measurement process. It has been possible to increase the range of $R$ by more than 50\%. Since the 
error in $\Delta E$ grows roughly proportional to $R^2$, this would be very 
hard to achieve by increasing computing power alone. Compared to \cite{Pushan2}, 
the errors on $E_1$ and $\Delta E_{10}$ have been reduced 
by a factor between 2 and 3 in the overlapping range of $R$. This would also be very hard to achieve by increasing 
statistics alone. Both of these observations clearly demonstrate the superiority of the new measurement process 
over the older one. Moreover using the older method we did not have a signal for the second excited state whereas 
we do have a reasonable measurement now at least for one $\beta$ value.

Our results seem to indicate that in this scheme, with our basis, upto the values of $R$ we have considered, one 
needs temporal extents of about 1 fm to make sure that the corrections due to the higher states 
are well under control. Wherever this criterion has been met, we have seen that the data follows 
the NG curve. 

The corrected data sets for different $\beta$ values seem to fall on top of each other indicating 
that there is very little effect due to finite lattice spacing. As such it seems to be much more 
important to go to larger temporal extents than finer lattices.

Unfortunately our data is still not good enough to distinguish between NG, LW and PS type string 
theories as that would require a sensitivity at the level of $R^{-5}$. Deviations from the NG predictions 
at higher orders have recently been reported in \cite{gliozzi} for gauge duals of random percolation problems.

This was our first attempt to show how to combine the two powerful techniques of multi-level error 
reduction and sophisticated sources for Wilson loops. Using this we have been able to go to 
much larger values of $R$ and $T$ than was possible before. 
For ruling out different string models, the error bars will have to be reduced by at least an order of 
magnitude. That at the moment looks like a project for the future. 

\vspace*{-2mm}\section{Acknowledgments}

The simulations were distributed over three different computational facilities, namely the computing resources
of the Westf\"alische Wilhelms-Universit\"at M\"unster (organized with the condor system \cite{condor}),
where this study was started, the teraflop Linux cluster KABRU at IMSc, Chennai and the Linux cluster 
LC2 at the ZDV of the
Johannes Gutenberg-Universit\"at Mainz.
We are indebted to the institutes for these facilities.
We would also like to thank P. Weisz and H. Wittig for useful discussions on finite size effects for Wilson loops.
B.B. is funded by the DFG via the SFB 443.

\end{document}